\def\swo{Sb$_2$WO$_6$}
\def\etal{{\it et al. }}
\def \tcr{\textcolor{black}}

\documentclass[preprint,5p,twocolumn]{elsarticle}
\usepackage{color}
\usepackage{graphicx}
\usepackage{epsfig}
\usepackage{caption}
\usepackage{subfigure}
\usepackage{amsmath}

\usepackage[english]{babel}
\usepackage[usenames,dvipsnames]{xcolor}
\definecolor{cream}{RGB}{222,217,201}

\begin{document}

\begin{frontmatter}

\title{Combined theoretical and experimental study of the electronic and optical property of Sb$_2$WO$_6$}

\author{Devdas Karmakar \fnref{fn1}}
\author{Sujoy Datta \fnref{fn1}} \ead{sujoydatta13@gmail.com}
\author{Debnarayan Jana \corref{cor1}} \ead{djphy@caluniv.ac.in}

\address{{Department of Physics, University of Calcutta, 92 A P C Road, Kolkata 700009, India.}}
\fntext[fn1]{Both contributed equally.}
\cortext[cor1]{Corresponding Author.}

\begin{abstract}
Both theoretical and experimental analysis are carried out to understand the \tcr{physical properties} of the fascinating electronic and optical properties of antimony tungstate (\swo). The nanosized ($\sim 40-80~nm$) material is produced using hydrothermal method followed by the SEM and XRD analysis to find the structural properties. \tcr{The present calculations using PBEsol and PBE approximations for exchange-correlation potential are compared with the experimental structural parameters and in the case of the calculations using PBEsol approach the predicted crystal parameters and simulated XRD pattern are in excellent agreement with experimental results. The experimental absorption spectra measured in the ultraviolet-visible range give the bandgap of $2.42 ~eV$, while the most intense peak of photoluminescence spectra is found at $468nm~(2.65 ~eV)$. Using density functional theory (DFT) technique, the band structure and density of states for \swo~ are calculated and the calculated bandgap of $2.62 ~eV$ is in agreement with the experimental finding.} From theoretical partial density of states calculations we identify that the bandgap is formed between the O$-2p_y$ orbitals bonded with Sb at valence band maxima and the W$-5d_{x^2-y^2}$ orbital at conduction band minima. The atomic level transitions responsible for the peaks of absorption spectra and photoluminescence spectra are identified as well by the means of DFT calculations. \tcr{Following the matching of theoretical and experimental observations, the calculations of optical properties reveal the plasma frequency to be equal to $13.36 ~eV$.}  
\end{abstract}

\end{frontmatter}

\section{Introduction}
Within the semiconductor family oxides occupy a special place due to their exotic physical and chemical properties. The vibrancy of their characteristics is exhibited through Mott insulation \cite{watanabe2010},  superconductivity \cite{shen1995}, a high dielectric constant \cite{robertson2005,homes2001}, and excellent electrical, optical, as well as electrochromic characteristics \cite{buch2016}. The partially filled d-shells of transition metals play the key role in producing such different novel fascinating properties whereas, the hybridisation with oxygen p-electrons tunes it further. Tungsten trioxide (WO$_3$) is such an example of semiconducting metal oxide exhibiting electrochromic property and various applications in field of gas sensing and other environmental remedies \cite{buch2016}.

\tcr{Metalloids, on the other hand, are p-elements in the Periodic Table often forming III-V semiconductors (e.g., GaAs, InSb).} However, they often form oxides. Among others, oxides of Antimony  also exhibit semiconductor properties, found in three distinct phases; antimony trioxide (Sb$_2$O$_3$), antimony tetroxide (Sb$_2$O$_4$), and antimony pentoxide (Sb$_2$O$_5$) \cite{chin2010}. 

While oxides of both tungsten and antimony are promising materials for application, their bandgaps are wider and fall under blue to ultraviolet (UV) region, e.g., the bandgap of WO$_3$ is of the order of $2.6-3.4 ~eV$  \cite{gonzalez2010,deb1973} and of Sb$_2$O$_3$ is $3.3 ~eV$ \cite{deng2006}. Interestingly, though most of the tungstates are widegap materials (Ag$_2$WO$_4: 3.15 ~eV$, Bi$_2$WO$_6: 2.8 ~eV$), Sb$_2$WO$_6$ is a relatively narrower bandgap candidate falling under visible region \cite{zhu2017,zhang2011}.

Semiconductors with bandgap falling in visible range are of particular interest in photocatalytic application and optoelectronic device designing.\tcr{Therefore, antimony tungstate, \swo, has attracted significant attention over recent decades.} \swo ~is a member of the Aurivillius family represented by the general equation [Sb$_2$O$_2$][A$_{m-1}$B$_m$O$_{3m+1}$] of unique layered structure by perovskite slabs, where, the A and B are transition metal atoms with $6$ and $12$ coordination, respectively \cite{ling1996,castro1994,castro1995,castro1997}. Here m is the number of consecutive perovskite layers. Bi$_2$WO$_6$ is one of the most renowned members of this family which is mostly used for the photocatalytic application for degradation of hazardous dyes in aqueous medium and hydrogen evolution by the decomposition of water \cite{yang2016,zhang2016,chen2018}. \swo, which is exhibiting its merit in similar field already, however, a detailed analysis on its characteristic features using coherent theoretical and experimental analysis is in due course. Theoretical understanding of the underlying mechanism governing the experimentally observed features of \swo ~should be helpful for further tuning of its properties which is yet to be explored. This study is an attempt to bridge the gap.

Here, we report a simple hydrothermal process of synthesizing \swo, followed by light absorption analysis in ultra-violet to visible \tcr{(UV-Vis)} region and photoluminescence spectral analysis. \tcr{To understand the mechanism behind these physical properties, the density functional theoritical study has been carried out.}





\section{Methodology}
\subsection{{Experimental Details}}
\subsubsection{Materials}
All the chemicals are of analytical grade and used without further purification. Antimony Chloride (SbCl$_3$) is purchased from Merck with $99.0\%$ purity. Sodium Tungstate (Na$_2$WO$_4$) is from Spectrochem Pvt. Ltd (India) with purity $99.0\%$. Deionised water (DI) is used for the synthesis purpose.

\subsubsection{Synthesis}
Antimony Tungstate nanostructures are prepared using antimony chloride (SbCl$_3$) and sodium tungstate (Na$_2$WO$_4$)  by the usual hydrothermal process \cite{rabenau1985}. Being a low cost, simple and easily controllable process, hydrothermal method is used for pure as well as doped and heterostructure sample preparation. The morphology and the average size of the nano particles can be controlled via tuning the temperature, time, sample concentration and pH of the solution.

Sodium tungstate and antimony chloride are dissolved in $22 ~ml$ DI water separately in $100 ~ml$ beakers followed by $15$ minutes sonication and $30$ minutes vigorous stirring. Sodium tungstate solution is then poured drop wise to the SbCl$_3$ solution and kept for stirring during $1$ hr to get a homogeneous mixture. After that the solution is transferred to the $55 ~ml$ teflon lined chamber in a steel autoclave at a temperature of $180^oC$ for $12$ hours. The solution is allowed to cool naturally and centrifuged several times with water-ethanol mixture to remove the unreacted salts. Finally, the product is dried at $70^oC$ for overnight and collected by mortaring.


\subsubsection{Characterization}
Structural characterization of the samples is carried out by using X-ray powder diffraction (XRD) measurements recorded using X-ray diffractometer (Model: Bruker AXS D8 Advanced) operating at 40 kV and $40 ~mA$ with Cu-K$_\alpha$ radiation of $1.5406 \AA$. XRD spectra are taken within $2\theta$ range of $15^o-70^o$ using a scanning rate of $0.003^o s^{-1}$. The surface morphology of \swo ~is studied by ZESIS EVO 18 scanning electron microscopy (SEM). The UV-Vis spectrum is recorded in the absorbance mode in the wavelength range $200 ~nm$ to $800 ~nm$ by using the instrument Shimadzu UV-1800. Photoluminescence (PL) spectra of the samples are taken by using Horiba FL 1000 fluorescence spectrometer using $310 ~nm$ excitation.

\subsection{Theoretical and Computational Methodology}
For density functional calculations we use plane wave basis functional techniques as implemented in the Quantum Espresso (QE) code \cite{QE}. \tcr{The ground state configurations are taken as: Sb$ \rightarrow 4d^{10}5s^25p^3$, W$\rightarrow 5s^25p^65d^46s^2$, O$\rightarrow2s^22p^4$.} The structural relaxation is carried out to find energetically optimized cell using variable cell structural relaxation technique. Pseudopotentials appropriate for projected augmented wave (PAW) basis is used. We use Perdew-Burke-Ernzerhof (PBE) and the corrected form of PBE for solids (PBEsol) exchange-correlation potentials for structural analysis \cite{PBE,PBEsol}. The structural optimization is done with the maximum force of $10^{-5} ~Ry./atom$ and pressure thresholds of $10^{-5} ~kbar/cell$. Fine  $6\times 7\times 4$ k-point grids ($\langle 0.2/\AA$) is used and charge densities and energies are converged to $10^{-8}~ Ry$. We  set kinetic-energy cut-off of $55 ~Ry$ and $500 ~Ry$ for the wavefunctions and charge densities, respectively. The simulated X-ray diffraction is produced using Vesta package \cite{vesta}

As PBE or PBEsol approximations underestimate the bandgaps, so, we use Heyd$-$Scuseria$-$Ernzerhof (HSE06) screened hybrid functional for electronic and optical properties calculations \cite{HSE06}. The  optimized norm-conserving Vanderbilt (ONCV) pseudopotentials are used for HSE calculations. The HSE06 band structure and density of states (DOS) are extracted using Wannier90 package \cite{wannier90}. For DOS calculation, adaptive smearing is utilised using denser $25\times25\times25$ k-space grid.

For optical properties predictions, we calculate the complex dielectric tensor using random phase approximation (RPA).
\small
\begin{align}
\epsilon_{\alpha\beta}&(\omega)= 1+\frac{4 \pi e^2}{\Omega N_{\textbf{k}} m^2}\sum\limits_{n,n'}\sum\limits_{\textbf{k}} 
\frac{\langle u_{\textbf{k},n'}\vert\hat{\textbf{p}}_{\alpha}\vert u_{\textbf{k},n}\rangle 
	\langle u_{\textbf{k},n}\vert\hat{\textbf{p}}_{\beta}^{\dagger} \vert u_{\textbf{k},n'}\rangle}
{(E_{\textbf{k},n'}-E_{\textbf{k},n})^2} . \nonumber\\
&\left[\frac{f(E_{\textbf{k},n})}{E_{\textbf{k},n'}-E_{\textbf{k},n}+(\hbar\omega+i\hbar\Gamma)} +
\frac{f(E_{\textbf{k},n})}{E_{\textbf{k},n'}-E_{\textbf{k},n}-(\hbar\omega+i\hbar\Gamma)}\right] 
\end{align}
\normalsize
Here, $\hat{\textbf{p}}$ is the momentum operator, $\vert u_{\textbf{k},n}\rangle$ is the $n$-th state at $\textbf{k}$ point corresponding to the energy $E_{\textbf{k},n}$, $\omega$ is the frequency of the incident photon, $\Omega$ is the volume of the unit cell and $N_k$ is the number density of the charge carrier.
Since no excited-state can have infinite lifetime, we introduce small inter-smearing $\Gamma$  ($0.2$) in order to incorporate intrinsic broadening to all excited states. We calculate the imaginary part of the dielectric function $\epsilon^{(i)}_{\alpha\beta}$ first, and, then the real part $\epsilon^{(r)}_{\alpha\beta}$ using the Kramers-Kronig relation.

\tcr{From} $\epsilon^{(r/i)}_{\alpha\beta}$ the optical conductivity, refractive index and absorption coefficients can be found as \cite{wooten}:
\begin{align}
&\text{Dielectric tensor:  } \epsilon_{\alpha\beta}=\epsilon_{\alpha\beta}^{(r)}+ i \epsilon^{(i)}_{\alpha\beta} \\ 
&\text{Optical Conductivity:  } Re [\sigma_{\alpha\beta} (\omega)]= \frac{\omega}{4\pi}\epsilon^{(i)}_{\alpha\beta}(\omega) \label{eq_cond}\\
&\text{Complex Refractive Index:  } \mu_{\alpha\alpha}=n^+_{\alpha\alpha}+i n^-_{\alpha\alpha}\\ 
&\text{Absorption Coefficient: } A_{\alpha\alpha}(\omega)=\frac{2\omega n^-_{\alpha\alpha}(\omega)}{c} \label{eq_abs}\\ 
& \text{where, } n^{\pm}_{\alpha\alpha}(\omega)= \sqrt{\frac{|\epsilon_{\alpha\alpha}(\omega)| \pm \epsilon^{(r)}_{\alpha\alpha}(\omega)}{2}} \nonumber
\end{align}

\section{Results and Discussions}
\subsection{Structural Properties}
\subsubsection{SEM Image}
\begin{figure}[]
	\begin{center}
		\includegraphics[scale=0.83]{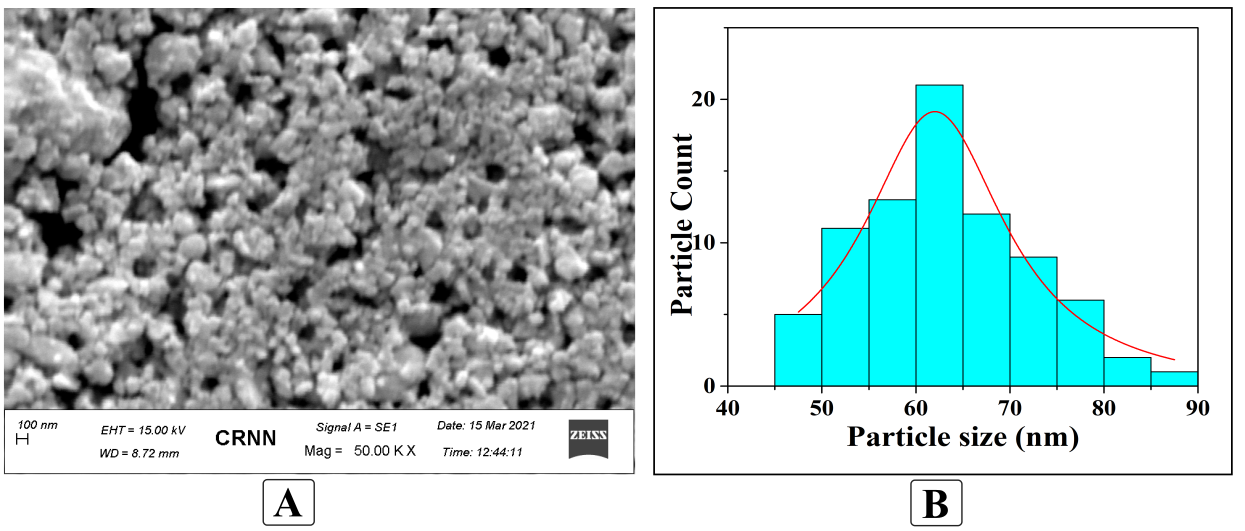}
		\caption{\label{SEM} (A) SEM image of the \swo ~nanoparticles; (B) Particle size distribution obtained from the SEM image and fitted with Gaussian.}
	\end{center}
\end{figure}

The surface morphology of the synthesized sample is studied using the SEM image. Fig.\ref{SEM}(A) exhibits the shape and sizes of the sample particles as prepared. It is clear that the particles are very small spherical balls and are almost uniform in shape and the size of those fall within nanoregion. The distribution of the measured particle sizes of randomly selected $80$ particles is presented by an histogram as shown in the Fig.\ref{SEM}(B). The sample particles are found to be of $40-80~nm$ size, while, most probable particle size obtained by Gaussian fitting as $62 ~nm$. Due to the small size of the particles, quantum confinement may give rise to the blue shifting of the bandgap.

\subsubsection{X-ray Diffraction Analysis}
\begin{figure*}[]
	\begin{center}
		\includegraphics[scale=0.97]{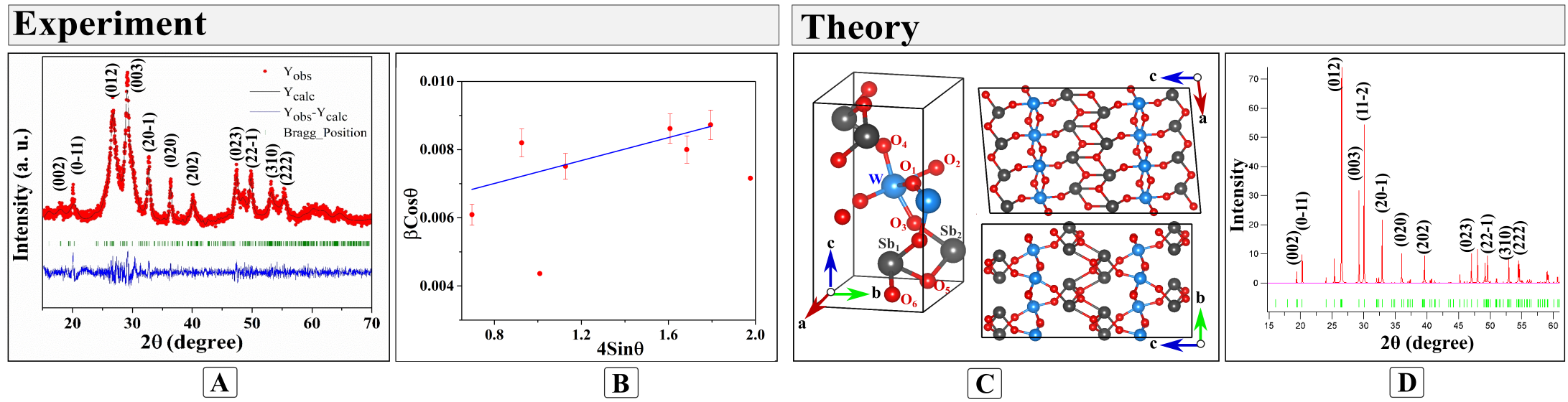}
		\caption{\label{strc} (A) The profile matching of XRD pattern of the \swo ~sample; (B) $\beta cos\theta$ versus $4 sin\theta$ plot (W-H plot) and the straight line represents the average fitting; (C) Crystal structure and (D) Simulated X-ray diffraction pattern for \swo.  }
	\end{center}
\end{figure*}

XRD technique is an effective tool to determine the phase, crystallinity and purity of the samples prepared. Fig.\ref{strc}(A) shows the XRD pattern of the synthesized \swo~ nanoparticles. The diffraction peaks at $2\theta$ values of $20.1^o, 27.0^o, 29.1^o, 32.9^o, 36.5^o, 40.3^o, 47.6^o, 49.9^o, 53.3^o$ and $55.5^o$ correspond to the crystal planes with (hkl) values as shown in the Fig.\ref{strc}(A) \cite{ding2018}. These peaks correspond to the triclinic phase of \swo~ [JCPDS card no. 47-1680] \cite{chen2018}. The absence of no other peak confirms the pure phase formation of \swo~ crystal.

\begin{table*}[]
	\centering
	\resizebox{\textwidth}{!}{%
		\begin{tabular}{cccccccc}
			\hline \hline
			\multicolumn{8}{c}{\textbf{Lattice Parameters}}                                \\ \hline
			& \textbf{Formula Units} & \textbf{a (\AA)} & \textbf{b (\AA)} & \textbf{c (\AA)}  & \textbf{$\alpha$} & \textbf{$\beta$}  & \textbf{$\gamma$} \\ \hline
			\textbf{Theo. (PBE)} & 2 & 5.641 &  5.087 &  9.267 &  90.00 & 95.11 & 90.00 \\
			\textbf{Theo. (PBEsol)} & 2 & 5.553 & 4.983 & 9.192 & 90.00 & 95.41 & 90.00 \\
			\textbf{Expt.} & 16 & 11.13 & 9.89 & 18.50 & 90.50 & 96.38 & 90.38 \\
			\textbf{Expt. \cite{castro1994}} & 2 & 5.554 & 4.941 & 9.209 & 90.05 & 96.98 & 90.20\\
			\textbf{Expt. \cite{ling1996}} & 16 & 11.132 & 9.896 & 18.482 & 90.20 & 96.87 & 90.21\\ \hline
			\multicolumn{8}{c}{\textbf{Band Gap (eV)}}                                     \\ \hline
			\textbf{PBE} & \textbf{PBEsol}        & \textbf{HSE06}   & \textbf{Expt.}   & \textbf{Theo. \cite{huang2020}} & \textbf{Expt. \cite{rafiq2019}} & \textbf{Expt. \cite{chen2018}} & \textbf{Expt. \cite{yang2016}}  \\
			1.78                    & 1.35 & 2.62  & 2.42  &  2.91     &  2.30     &   2.46    &  2.17$-$2.63     \\ \hline \hline
		\end{tabular}%
	}
	\caption{	\label{tab_latpar} Lattice parameters and bandgap of \swo~ calculated using DFT methods compared with the experimental values obtained. Experimental lattice parameters of this study and that by Ling. \etal \cite{ling1996} are for $2\times2\times2$ supercell.}
\end{table*}

\tcr{The broadening of the XRD pattern arises due to two reasons: finite size of the crystals and the strains developed in them.} This strain arises due to formation of the defect in the crystal structure during synthesis as well as effect from grain boundaries. The Williamson-Hall (W-H) plot is a fruitful method to separate the broadening due to the above-mentioned two causes and also to determine the grain size of the crystal \cite{cullity1956}. Accordingly, the total broadening $\beta_{tot}$ of XRD peaks is equal to the sum of the  broadening due to grain size ($\beta_{size}$) and microstrain ($\beta_{strain}$) developed in the crystal structure that can be written as 
\begin{equation}
	\beta_{tot}=\beta_{size}+\beta_{strain}
\end{equation}

Using the crystalline size $D$, the microstrain $\epsilon$ and the shape factor $k=0.9$ for spherical particles, the broadening due to crystal size and microstrain canbe written as: Thus, the total strain:
\begin{align}
\beta_{size}= \frac{k\lambda}{D cos \theta} ~&;~ \beta_{strain} = 4\epsilon ~tan \theta  \nonumber\\
	\beta_{tot}= \frac{k\lambda}{D cos \theta} + 4\epsilon ~tan \theta
	&\Rightarrow \beta_{tot}~ cos \theta = \frac{k\lambda}{D} + 4\epsilon ~sin \theta
\end{align}

In Fig.\ref{strc}(B), we present the W-H plot as calculated from the XRD pattern. From straight line fitting of the major peaks of the XRD pattern, we get the slope of the straight line is $0.00168 \pm 0.00091$. This is the equal to the strain developed in the crystal. While the intercept of the straight line is $0.00568 \pm 0.000668$ and the corresponding average  grain size is $24.59 ~nm$. 

Using the initial input parameters provided by  Ling \etal \cite{ling1996}, the Rietveld analysis is done using FullProf software. The calculated lattice parameters as presented in Table~\ref{tab_latpar} is similar to the previously reported values \cite{ling1996,castro1994}. The small deviation of angles $\alpha$ and $\gamma$ from $90^o$ in all of these experiments suggest a triclinic phase of \swo. 

\subsubsection{Theoretical Analysis}
\swo ~is a crystal \tcr{which follows} perovskite structure of layered Aurivillius phase \cite{ling1996,castro1994}. In Fig. \ref{strc}(C) the unit cell of \swo ~is depicted. Each unit cell contains two formula \tcr{units} totalling $18$ atoms. Sublayers are formed by WO$_4$ (let say, A) and Sb$_2$O$_2$ (B) in alternating ABAB$...$ arrangement \cite{ramirez1997}. \tcr{The $2\times2\times2$ supercell of the structure clearly indicates the layered structure as seen in Fig.\ref{strc}(D).} The arrangement of the Sb$_2$O$_2$ sublayer follows a zigzag pattern resulting a lower symmetry of the crystal.

The lattice parameters found by our DFT study using PBE exchange-correlation functional are $a=5.641 \AA$,  $b=5.087 \AA$ and $c= 9.267\AA$, while, the angles are $90.00 ^o$, $95.11 ^o$ and  $90.00 ^o$. This is comparable with the experimental finding of Castro \etal ($5.554 \AA, 4.941 \AA,  9.209\AA$; $ 90.05 ^o, 96.98 ^o, 90.20^o$) \cite{castro1994}. The little overestimation of bond-lengths is an usual feature of PBE functional as it underbound atoms \cite{PBEsol}. For a large set of semiconductors the mean of overestimation is calculated as $1.502 \%$ \cite{datta2020rev} and  in \swo ~the mismatch is just $2.96 \%, 0.64 \%, 1.57\%$ for $a,b,c$, respectively. The W atom forms six W-O bonds among which four are almost planar in nature having O-W-O bond angles $100.67^o, 87.11^o, 84.03^o, 88.17 ^o$. The on plane W$-$O bond lengths are $1.81\AA, 2.12 \AA, 2.16 \AA, 1.80 \AA$ and the two out-of-the-plane bond lengths are $1.89 \AA$. So, we see that there is inherent asymmetry within the WO$_4$ sublayer whereas the inter-sublayer (Sb$_2$O$_2$-WO$_4$) region is maintaining a regularity.

On the other hand, PBEsol predicted structural parameters ($5.553 \AA, 4.983\AA, 9.192\AA,  90.00^o, 95.41^o, 90.00^o$) are in excellent agreement with the experimental values when compared with PBE functional. This is because the slow varying density approximation is more accurate for any solid and in the PBEsol functional that is restored \cite{PBEsol}.

\tcr{Though the lattice parameters calculated using PBEsol are almost similar to the experimental parameters found by Castro \etal, the mismatch of symmetry group is evident. While the experimental observation shows that \swo~ follows triclinic $P1$ space group symmetry, the DFT calculations using both PBE and PBEsol indicate it as monoclinic ($P2_1$) crystal. The $\alpha$ and $\gamma$ angles are found to be $90^o$ using DFT calculations, whereas, the experimentally observed values deviate a little from $90^o$ (see, Table~\ref{tab_latpar}). The deviation of the angles from a perfect $90^o$ is responsible for the triclinic nature. However, Ling \etal have experimentally predicted the underlying monoclinic structure of \swo~ crystal \cite{ling1996}. For experimental observations, such variation is not rare as finite size effect is inevitable in experiments. For example, the lattice angle $\beta$ of iron tungstate FeWO$_4$, which is found in $P2/a$ symmetry, may vary from $90.091^o$ to $90.8^o$ \cite{rajagopal2010,yu1993,zhou2009}.} Interestingly, the enantiomorphic nature (non-centrosymmetric) is predicted by both theoretical and experimental techniques.

The theoretically calculated XRD pattern is presented in Fig.\ref{strc}(D). The calculated peaks match well with the experimental observation. Unlike the experimental case, the calculated peaks are sharp and one can distinctly identify many of them. This is because the theoretical calculation is done on bulk material free from any size effect. The size effect makes the experimental XRD peaks blunt and, sometimes, less distinctly observable. \tcr{Due to the broadening of the XRD peaks, the peak corresponding $(11-2)$ is not visible in the experimental XRD. However, by deconvoluting the experimental XRD spectra between $2\theta$ range $28^o -32^o$ , two peaks can be located. This is shown in Fig.\ref{XRD} of Appendix. The peak at $29.25^o$ is due to $(003)$ plane and peak at $30.11^o$ is due to $(11-2)$ plane.}

\subsection{Electronic and Optical Properties}

\subsubsection{Ultraviolet-Visible Spectra}
The optical property of the material is studied by UV-Vis absorbance spectroscopy. The absorption of photons energizes the electron to jump from valence band (VB) to conduction band (CB), so, the absorption spectra provides the understanding of the electronic properties of the material concerned.

\begin{figure*}[h]
	\begin{center}
		\includegraphics[scale=0.75]{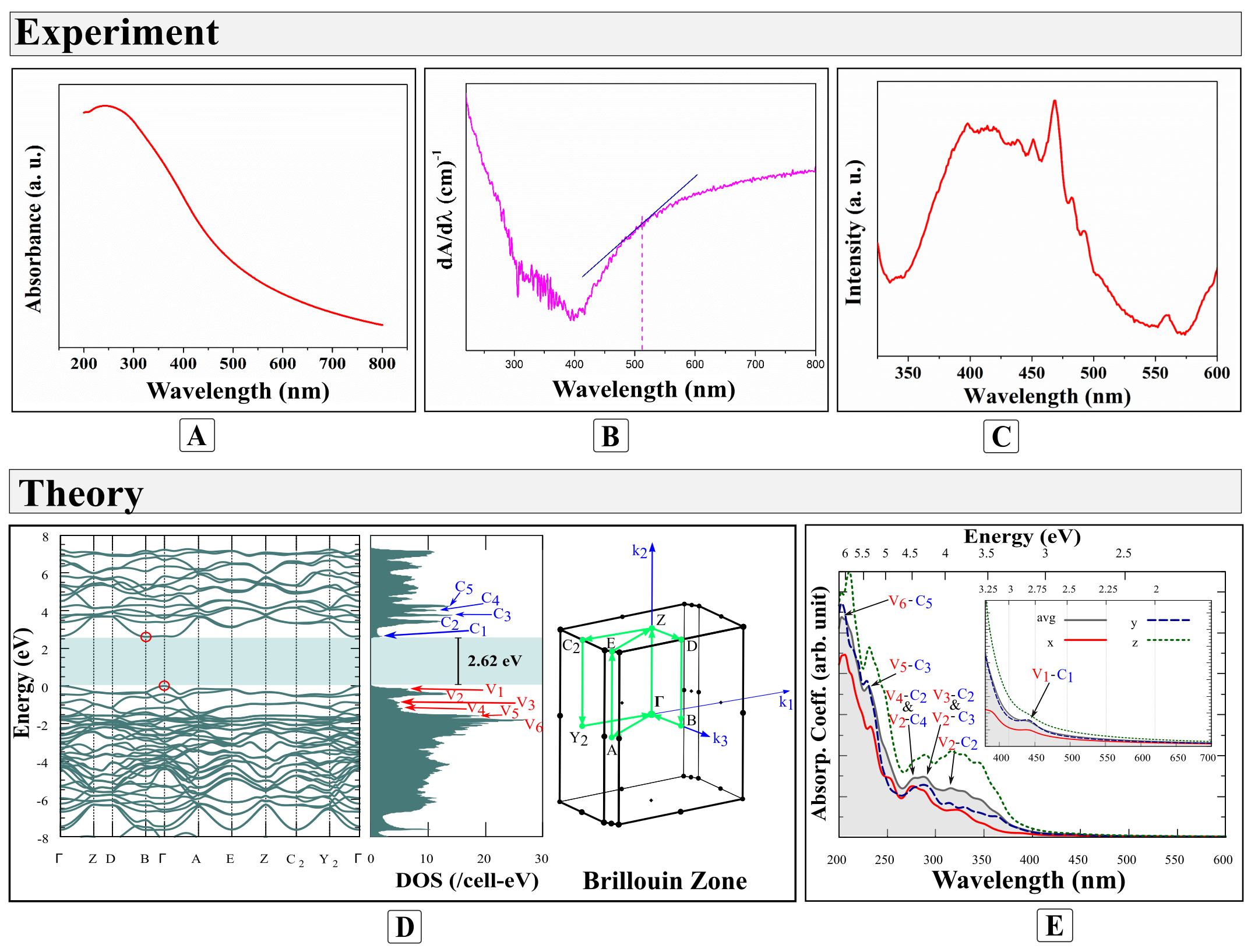}
		\caption{\label{elec} (A )UV-Vis absorbance spectra in water solution; (B) The $\frac{dA}{d\lambda}$ versus wavelength plot; (C) Photoluminescence spectra of the antimony tungstate; (D) Energy band structure and \tcr{density of states} and (E) Absorption Coefficient calculated using HSE06 hybrid functional for \swo.  }
	\end{center}
\end{figure*}

The observed UV-Vis spectra of the sample is shown in Fig.\ref{elec}(A).
We can see from the figure there is a prominent absorbance ($A$) of the material in the visible region.
Following an almost flat region in $\sim 200-208~nm$ range, the absorbance start to increase and there is a peak observed at $242-245 ~nm$. Beyond that the intensity starts to fall rapidly till $450~nm$ followed by a rather moderate fall till $800~nm$.

Now, the absorbance is observed for a solution of \swo, so, the effect of the interaction between water and the sample surface on the spectra is unavoidable. Due to the interaction between water and the sample surface there is a little flatness in the absorbance spectra and very few prominent peaks are noticed. Similar nature of absorbance is observed in earlier works as well \cite{ding2018}. There is no sharp fall in absorbance near the bandgap energy attributing from this interaction as well. So, to detect the region of maximum change of absorbance we plot the $\frac{dA}{d\lambda}$ versus wavelength $\lambda$ as shown in Fig.\ref{elec}(B). The knee point of $\frac{dA}{d\lambda}$, where a straight-line fitting is done, can be identified as the point where the absorption just start increasing. The wavelength of this point should correspond the bandgap energy which is estimated as $2.42 ~eV$.

\subsubsection{Photoluminescence Spectra}
In general, the recombination of photo-generated charge carriers can release energy for the emission of photoluminescence. The higher the PL intensity, higher the recombination of charge carriers and more the sample acts like fluorescent material. Fig.\ref{elec}(C) shows the photoluminescence spectra of antimony tungstate with the excitation of 310 nm. There is a prominent part of the emission in the visible region.

\tcr{The most intense peak at $468 ~nm$ is attributed to the intrinsic luminescence of \swo~ \cite{rafiq2019}.} The approximate bandgap energy as calculated from this peak corresponding to the PL spectra is $2.65 ~eV$. From the theoretical partial DOS (PDOS) calculation discussed in the next subsection, we can identify that the peak originates from the electron transitions between the hybridized Sb$-5s$ orbital and O$-2p$ orbital  lying in the VB and W$-5d_{x^2-y^2}$ orbital in the CB \cite{xiao2008,li2014}. The peak at $438 ~nm$ may be formed due to the defect of the metal atoms and the oxygen vacancies produced during the crystal growth \cite{xiao2008}.

\begin{figure*}[h]
	\begin{center}
		\includegraphics[scale=1.0]{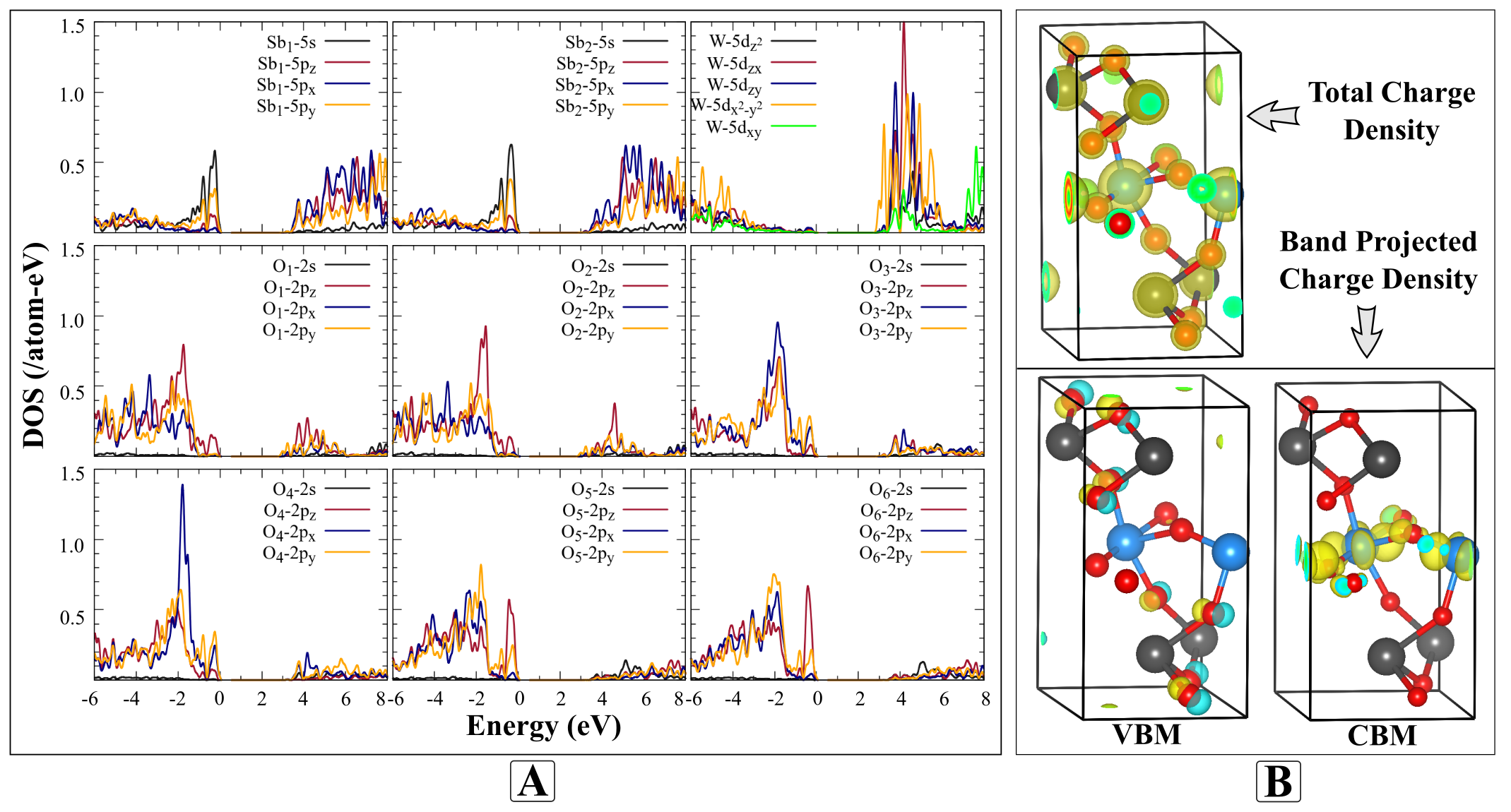}
		\caption{\label{pdos} (A) Atomic orbital projected DOS  of \swo; (B) Total charge density and the charge densities for VBM and CBM at the $\Gamma$ and $B$ points, respectively.}
	\end{center}
\end{figure*}

\subsubsection{Band Structure and Density of State}
The electronic structural analysis provides the understanding of the optical performance of materials, hence we investigate the energy band structure and DOS of the optimised structure. Both the PBE and PBEsol generalised gradient approximations (GGA) perform well for structural analysis, however, they underestimate the \tcr{bandgaps} of the semiconductors and insulators. The derivative discontinuity effect can not be properly addressed within such non-empirical semi-local approximated schemes \cite{perdew1982,datta2020rev}. The Hartree-Fock (H-F) approximation is a non-local exchange scheme and self-interaction free, so, in hybrid functional schemes a portion of H-F exchange is added. The HSE06 screened hybrid method is one of the best performing technique for proper bandgap prediction and is computationally less demanding than the density functional perturbative theoretical approximations ($GW$ based quasi-particle methods) \cite{HSE1}. We use HSE06 hybrid functional for electronic structural calculations as well as for optical property prediction \cite{HSE06}.   

In Fig.\ref{elec}(D), the energy band dispersion and DOS using HSE06 is depicted. \tcr{The coordinates of high symmetry points can be found in Table~\ref{tab_BZ} of Appendix. The bandgap is estimated as $2.62 ~eV$. The valence band maxima (VBM) is located at $\Gamma$ point and the conduction band minima (CBM) is at $B$ point, hence the bandgap is indirect in nature. From Table~\ref{tab_latpar}, we see that the bandgap is within the range of experimentally observed values. The energy band dispersion near the CBM is flat along the $\Gamma-B$ line.} Due to this flatness of the band there is a small peak of DOS at CBM denoted by $C_1$. There is a degeneracy of two bands along the $E-Z-C_2$ and $Z-D$ segments of irreducible Brillouin zone (BZ) in both of lowest lying conduction and highest lying valence band region. The almost separate two bands just below VBM is the reason of the sharp $V_2$ peak. The \tcr{overlap} of bands for a large region of BZ indicates strong hybridization. Similar degeneracy is observed for the next two bands in VB. It is interesting that the band degeneracies do not appear near the BZ centre $\Gamma$. While there are a series of distinct sharp peaks of DOS in CB denoted by $C_2,C_3,C_4 \cdots$, the DOS in VB is comprised of small peaks ($V_1, V_3, V_4, \cdots$) except the tall sharp peak $V_2$ discussed above.

The origin of the energy bands can be understood using the \tcr{atomic orbital projected DOS}. In Fig.\ref{pdos}(A) we present the PDOS for Sb, W and O atoms. If we look back to the crystal structure in Fig.\ref{strc}(C), we see that the Sb$_1$ and Sb$_2$ atoms produce bonding with O$_5$ and O$_6$. \tcr{The PDOS plots suggest hybridization of Sb$-5s$ with O$_5-2p_y$ and O$_6-2p_y$ near the VBM level, with a small contribution from Sb$-5p_y$, corresponding $V_1$ and $V_2$ peaks in Fig.\ref{elec}(D). The $V_5$ and $V_6$ peaks are originating from oxygen $2p$ levels; mainly the $2p_z$ orbital of O$_1$ and O$_2$, and, $2p_x$ orbital of O$_3$ and O$_4$ produce the sharpness. Especially O$_4-2p_x$ PDOS is distinctly sharp. CB is populated by W$-5d$ orbitals producing the spaghetti of sharp peaks of DOS. Near the CBM contribution is coming mostly from W$-5d_{x^2-y^2}$ orbital, with small contribution from the $2p$ orbitals of surrounded $O_1$ to $O_4$ atoms.}
\tcr{ Hence, the PDOS plot in Fig.\ref{pdos} demonstrates that O$-2p$ states are principal contributors to the upper portion of the valence band and W$-5d$ states give the main input in the lower portion of the valence band. On the other hand, the conduction band is dominated by unoccupied W$-5d$ states. It is worth mentioning that, the same feature regarding occupation of the valence band and conduction band regions by O$- 2p$ and W$- 5d$ states was found to be characteristic of other related tungstates with the common formula MWO$_4$ (M = Fe, Co, Cu, Zn, Cd) \cite{khyzhun2009,atuchin2016,rajagopal2010,khyzhun2013}.}

To be more specific about the atomic contribution at VBM and CBM, we calculate the band projected charge density at the VBM for $\Gamma$ point and at CBM for $B$ point. This plot is similar to the wannier orbital plot as the charge density is basically the squared value of the wavefunction. \tcr{The plots presented in Fig.\ref{pdos}(B) confirm that the bandgap is between the O$_5-2p_y$ and O$_6-2p_y$ at VBM and W$-5d_{x^2-y^2}$ at CBM. Small contributions from O$_1$ and O$_2$ at CBM is evident in the figure indicating the hybridization.} When we perform the wannierization to find maximally localised Wannier functions (MLWF) we find that the Wannier centres are shifted from the atom centres, which confirms the hybridization of orbitals.

\begin{table}[]
	\centering
	\begin{tabular}{cc}
		\hline \hline
	{\bf Atom} & {\bf Charge} \\ \hline
	Sb$_1$	& 1.82 \\
	Sb$_2$	& 1.83 \\
	W		& 2.43 \\
	O$_1$	& -0.87 \\
	O$_2$	& -0.89 \\
	O$_3$	& -1.03 \\
	O$_4$	& -1.03 \\
	O$_5$	& -1.14 \\
	O$_6$	& -1.13 \\ \hline
	\end{tabular}
\caption{	\label{tab_bader} Bader charge transfer.}
\end{table}

The atomic bonding nature can be qualitatively understood using Bader charge transfer analysis \cite{tang2009}. The calculated transferred charges for the atoms are presented in Table~\ref{tab_bader}. It shows that the polyhedra formed by W atom surrounded by O$_1$ to O$_4$ atoms has a net charge of $-1.75$ and Sb$_2$O$_2$ have a net charge of $1.75$. So, we can conclude that there is a net charge transfer from the Sb$_2$O$_2$ sublayer to WO$_4$ sublayer producing an ionic bonding between these two sublayers. The total charge density plot in Fig.\ref{pdos}(B) also indicates the ionic bonding nature as the charges around the atoms are almost spherically symmetric.  
 
So, we identify the origin of the bandgap as well as the nature of atomic bonding from the DFT study.
Also, the calculated gap comes in good agreement with that found from PL spectra. Motivated by such excellent agreement between the theoretical and experimental observations, we do further theoretical calculations on the optical properties.

\subsubsection{Absorption Coefficient}
As HSE06 functional can estimate the bandgap almost exactly, so, we further use it for the calculation of optical properties which are directly related to the electronic structure of materials. The jump of electron from a state to another is the mechanism behind the optical response; in metals both of inter-band and intra-band transitions are important, whereas, in gapped systems the optical response is  dominated by inter-band leap. Ideally, a photon with a minimum of bandgap energy is required to move an electron from VB to CB producing a hole in the VB. Photons with higher energy than bandgap initiates the transition to higher level of conduction band, hence, in optical response calculations a plenty of empty conduction bands are thus necessary to be taken into consideration. In VB region, the number of available electrons for the inter-band transitions is expressed in term of effective electron number (n$_{eff}$) as: 

\begin{equation}
n_{eff}(E_m) = \frac{2m}{Ne^2h^2}\int_0^{E_m}E . \epsilon^{(i)}_{\alpha\alpha}(E) dE;
\end{equation} 
Here, m, e and N are electronic mass, charge and number density. 
Calculation reveals that the electrons below $15 ~eV$ from VBM cannot contributing in the optical transitions. As a consequence, $0-15 ~eV$ range is a optimal choice of optical property calculations.

The absorption coefficient as calculated using Eq.\ref{eq_abs} is presented in Fig.\ref{elec}(E); on the shaded region of total absorption, the individual components are depicted as well. All the other optical properties are discussed in the following section, here we try to compare the experimental and theoretical absorption results side by side to get an insight on the underlying physical phenomena. 

Similar to the experimental observation, there is a small peak near $200 ~nm$. The sharp fall following that is also evident, however, in theoretical calculation the fall is stiffer than the experimental one. The fall goes on till $\sim 265 ~nm$ and thereafter a plateau is seen with almost flat top in $\sim 275-290 ~nm$ region. Following that, the fall is gradual and there is no absorption after $\sim 475 ~nm$. The calculated bandgap is $2.62 ~eV$ corresponding $473.22~nm$, so, there should not be any absorption beyond that. As, comparing to the higher energy region the absorption in visible light region ($\sim 380-700 ~nm$) is quite low, so,we magnify the plot in the inset for this region. The energy corresponding to the wavelength range is also depicted on the upper x-axis, in both of the plots.

Now, let us investigate the origin of the peaks in the absorption coefficient plots starting from lower energy towards higher energy (higher to lower wavelength) i.e., from right to left of Fig.\ref{elec}(E). The first distinct peak is at $\sim 2.8 ~eV$ (see, inset) corresponding to the $V_1-C_1$ transition. The transition between the first two sharp peaks $V_2-C_2$ is seen around $3.9 ~eV$. The next two peaks indicated in the figure are higher than the surroundings due to two transitions for each one. As there is a sharp rise of DOS from $V_4$ to $V_5$, the absorption coefficient rises sharply till $5.38 ~eV$ corresponding $V_5-C_3$ peak, and the next peak is near $200 ~nm$ ($\sim 6 ~eV$) is due to the $V_6-C_5$ transition.

Along with these indicated peaks, it is expected to have a distinct $V_2-C_5$ transition peak at $\sim 4.7 ~eV$, however, there is none. When minutely investigated, we note that the valley between $C_3-C_4$ and $V_2-V_3$ have a gap of $\sim 4.7 ~eV$ as well, so, we conclude that in the backdrop of this the expected $V_2-C_5$ is absent. 

We have discussed the similarities of the experimental and theoretical absorption spectra, however, we should note the difference of those as well. The DFT calculation is on the bulk, whereas, the experimental observation is on nano sized materials. The surface effect for small sized particles play significant role in optical absorption. Furthermore, there are interactions between water and the sample and the effect of the solvent water molecules makes the experimental spectra smoother. Though in theoretical calculation there should not be any absorption below the bandgap of the material, the scenario of experimental setup is quite different. We observe very low, but not tending to zero, absorption below the bandgap energy and the surface effect as well as the solvent environment is the reason behind such usual behaviour in experiments.

\subsubsection{Other Optical Properties}
We see that by proper choice of DFT method, one can find excellent agreement of the structural and electronic properties as well as can dig out the atomic level reasoning behind the experimentally observed photoluminescence and absorption profile. Such understanding let us move further to the theoretical studies on other optical properties. The calculation method is straight forward as discussed in the Methodology section.

\begin{figure}[h]
	\begin{center}
		\includegraphics[scale=0.6]{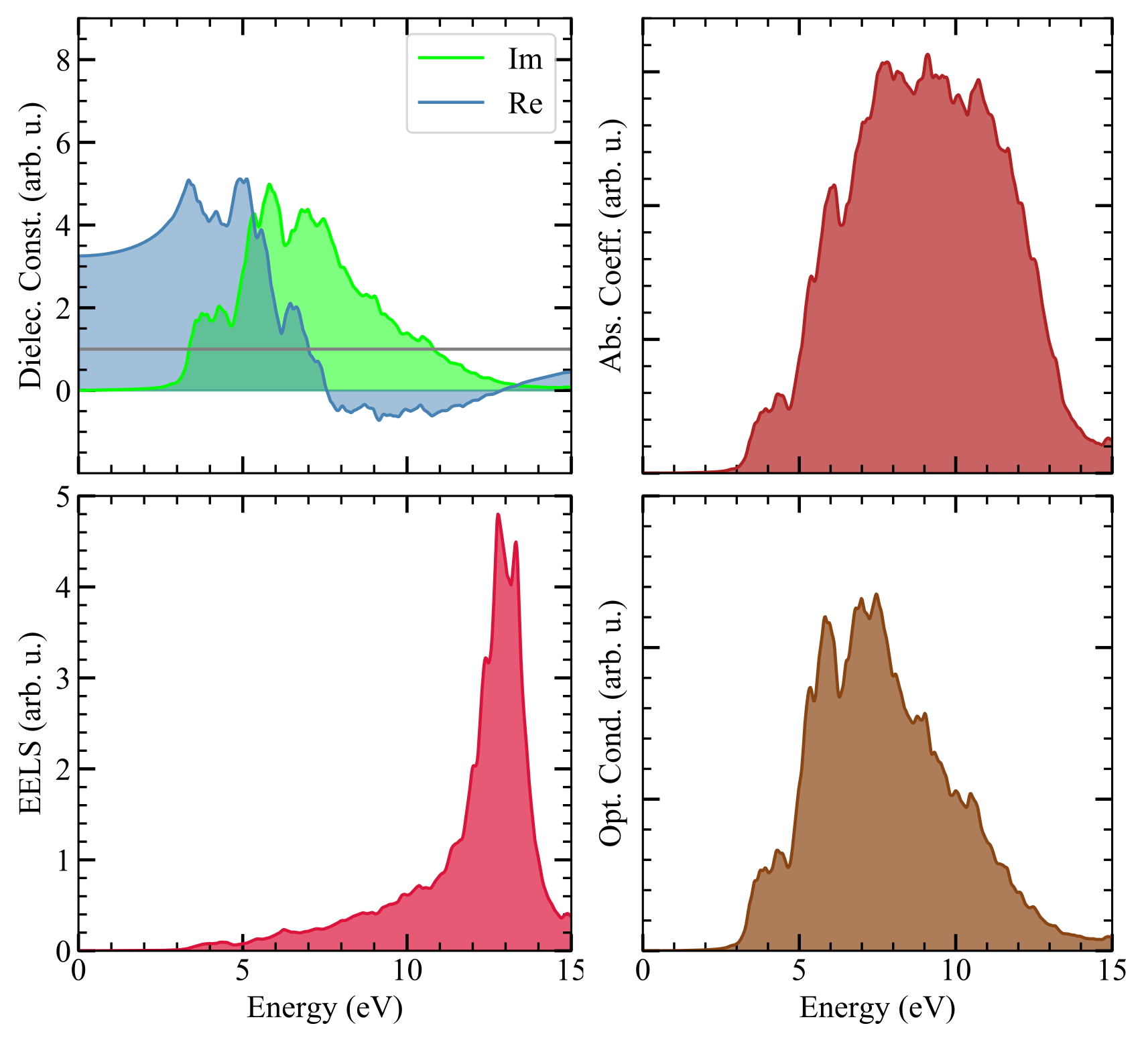}
		\caption{\label{opt} Theoretically calculated dielectric constant, electron energy loss spectrum, absorption coefficient and optical conductivity  of \swo.}
	\end{center}
\end{figure}

In Fig.\ref{opt} we plot the dielectric constant, electron energy loss spectrum (EELS), absorption coefficient ($A$) and optical conductivity ($\sigma$) of \swo. The imaginary part of dielectric constant $\epsilon^{(i)}$ remains zero till the bandgap energy. It then starts to rise rapidly till reaching a small flat peak region $~3.5 ~eV$ before it restarts its climb again. Near $5.5 ~eV$ the $\epsilon^{(r)}$ becomes higher than $\epsilon^{(i)}$.  The plasma frequency is defined as the frequency when $\epsilon^{(r)}$ crosses the zero axis going from negative to positive value while having $\epsilon^{(i)} \langle 1$ which is calculated as $13.36 ~eV$.

The plasma frequency can be readily verified from the distinct peak in EELS. The EELS is seen as a sharply peaked spectrum with very low value below $10 ~eV$. High value of plasma frequency is a signature character of most of the semiconductors, \swo~ is not an exception in that sense.

The same absorption coefficient presented in Fig.\ref{elec}(E) is now plotted for a wider range of energy, here, in this figure. We see that beyond the $6.2 ~eV$ energy corresponding to the $200 ~nm$, the absorption coefficient rises further and exhibits an almost bell shaped nature. The highest range of absorption is in $7.2-11 ~eV$ range. This is because of the highest range of DOS in VB, with peak at $V_6$ and fall $\sim-7 ~eV$ with respect to the VBM (see, Fig.\ref{elec}(D)).

The optical conductivity is directly related to the imaginary part of the dielectric constant through Eq.\ref{eq_cond}. In Fig.\ref{opt} such correspondence is readily visible. Now, absorption of photonic energy to produce electron-hole combinations is represented in absorption coefficient while the optical conductivity is the representative of the conducting property of the same electron-hole pairs. As larger number of electrons moves to CB through photon absorption, the carrier concentration also increases which reflects in higher optical conductivity. This signature is also evident from the plots. The conductivity jumps up just above the bandgap value ($2.62~eV$), and, even below $5~eV$ the optical conductivity is not very low. The flatness of the bands in the vicinity of VBM and CBM makes the conductivity profile almost flat for $\sim2.6-4.9 eV$. Beyond that the sharp increase follows the profile of absorption spectra.

\section{Conclusion}
We provide a \tcr{combined} theoretical and experimental study on \swo~ exhibiting its merit. The sample is produced using hydrothermal procedure. The XRD measurement proves its purity, whereas, the W-H plot and SEM measurements show that the particle size is of the order of $40-80 ~nm$. The peaks of the simulated XRD \tcr{pattern} for \swo~ match well with the experimental observation. The UV-Vis absorption spectra shows similar pattern as the DFT calculated one. \tcr{ From UV-Vis spectra the bandgap is found as $2.42 ~eV$, whereas, the bandgap calculated using HSE06 functional ($2.62 ~eV$) falls within the range found in experiments.} Thus with the help of energy band diagram and \tcr{atomic orbital} projected DOS we can identify the transitions responsible for the particular nature of the absorbance experienced experimentally. \tcr{The origin of the bandgap is identified as the energy difference between states occupied by the O$-2p_y$ orbitals bonded with the Sb atom at VBM, and, the W$-5d_{x^2-y^2}$ orbital at CBM. From the PDOS plot, the peak of PL spectra at $468~nm (2.65~eV)$ can be associated with the transition between hybridized Sb$-5s$, O$-2p$ at VB and W$-5d$ at CB.}  The significantly higher plasma frequency calculated for \swo~ ($13.36 ~eV$) is the character of semiconductors. So, our attempt of understanding the underlying physics of experimentally promising \swo~ paves the way for further tuning of its optical and electronic properties as well as establishes the importance of synchronization of experimental and theoretical works.  

\section*{Credit Author Statement}
The experimental characterisation and analysis are done by D. Karmakar and D. Jana and the theoretical analysis is done by S. Datta. With the contribution of D. Karmakar and D. Jana in experimental section, the manuscript is mainly developed by S. Datta. D. Karmakar and S. Datta contribute equally to this study.

\section*{References}

\bibliographystyle{elsarticle-num}

\appendix
\section*{Appendix}
\begin{figure}[h]
	\begin{center}
		\includegraphics[scale=1.5]{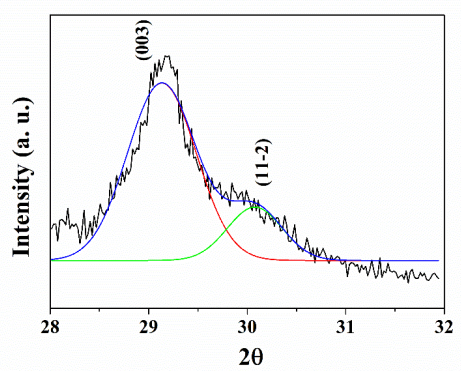}
		\caption{\label{XRD} Deconvoluted XRD spectra between $2\theta$ range $28^o -32^o$}
	\end{center}
\end{figure}

\begin{table}[]
	\centering
		\begin{tabular}{cccc}
			{\bf Labe}l    & $k_1$ & $k_2$ & $k_3$ \\\hline
			$\Gamma$ & 0.0   & 0.0   & 0.0   \\
			A        & -0.5  & 0.0   & 0.5   \\
			B        & 0.0   & 0.0   & 0.5   \\
			C$_2$    & -0.5  & 0.5   & 0.0   \\
			D        & 0.0   & 0.5   & 0.5   \\
			E        & -0.5  & 0.5   & 0.5   \\
			Y$_2$    & -0.5  & 0.0   & 0.0   \\
			Z        & 0.0   & 0.5   & 0.0  
		\end{tabular}%
	\caption{\label{tab_BZ} High symmetry points of Brillouin zone.}
\end{table}
\end{document}